\documentclass[
    ,final            
  ]
  {aipproc}

\layoutstyle{6x9}

\catcode`\@=11
\def\gsim{\ifmmode{\mathrel{\mathpalette\@versim>}}
    \else{$\mathrel{\mathpalette\@versim>}$}\fi}
\def\lsim{\ifmmode{\mathrel{\mathpalette\@versim<}}
    \else{$\mathrel{\mathpalette\@versim<}$}\fi}
\def\@versim#1#2{\lower 2.9truept \vbox{\baselineskip 0pt \lineskip
    0.5truept \ialign{$\m@th#1\hfil##\hfil$\crcr#2\crcr\sim\crcr}}}
\catcode`\@=12 

\def\kappaff{\kappa_{\rm ff}}
\def\Bnu{B_{\nu}}
\def\rr{{\cal R}}
\def\kb{k_{\rm B}}
\def\nscat{N_{\rm \gamma e}}
\def\brem{bremsstrahlung$\;\,$}
\def\Mbh{M_{\rm BH}}
\def\gef{g_{\rm eff}}
\def\eps{\epsilon}

\def\sigff{\sigma_{\rm ff}}
\def\sigtot{\sigma_{\rm t}}
\def\sigT{\sigma _{\rm T}}
\def\sigkn{\sigma_{\rm KN}}

\def\sigknnu{\sigkn (\nu)}

\def\Lx{L_{\rm X}}
\def\Luv{L_{\rm UV}}
\def\Lbh{L_{\rm BH}}
\def\Ledd{L_{\rm Edd}}
\def\LeddKN{\Ledd^{\rm es}}
\def\LeddKNff{\Ledd^{\rm es+ff}}

\def\jff{j_{\rm ff}}

\def\Tuv{T_{\rm UV}}

\def\fx{f_{\rm X}}
\def\fuv{f_{\rm UV}}
\def\fz{f_{\rm BH}}
\def\fznu{\fz (\nu)}

\def\Zi{Z_i}
\def\nel{n_{\rm e}}
\def\nei{n_{\rm i}}
\def\mpr{m_{\rm p}}
\def\mel{m_{\rm e}}
\def\nh{n_{\rm H}}
\def\nhe{n_{\rm He}}
\def\nt{n_{\rm t}}

\def\nuT{\nu_{\rm T}}
\def\nub{\nu_{\rm b}}
\def\nup{\nu_{\rm p}}

\def\qe{q_{\rm e}}

\begin{document}

\title{Free-free absorption effects on Eddington luminosity}

\author{L. Ciotti}{
  address={Astronomy Department, Bologna University, Italy}}
\author{J.P. Ostriker}{
  address={IoA, Cambridge, UK and Dept. of Astrophys. Sciences, Princeton 
              University, NJ, USA}}

\begin{abstract}
In standard treatments the Eddington luminosity is calculated by
assuming that the electron-photon cross section is well described by
the Thomson cross section which is gray (frequency independent). Here
we discuss some consequence of the introduction of free-free opacity
in the Eddington luminosity computation: in particular, due to the
dependence of \brem emission on the {\it square} of the gas density,
it follows that the associated absorption cross section increases {\it
linearly} with the gas density, so that in high density environments
Eddington luminosity is correspondingly reduced. We present a summary
of an ongoing exploration of the parameter space of the problem, and
we conclude that Eddington luminosity in high density environments can
be lowered by a factor of ten or more, making it considerably easier
for black holes to accelerate and eject ambient gas.
\end{abstract}

\maketitle


\section{Introduction}

The Eddington luminosity plays a fundamental role in our understanding
of accretion phenomena and, as a consequence, of the AGN physics and
evolution. In turn, it is also clear that AGNs played a central role
in galaxy formation, as testified by several empirical scaling
relations between the mass of the supermassive BHs at the center of
stellar spheroids (i.e., bulges and elliptical galaxies) and global
properties of the spheroids themselves.  In standard treatments
Eddington luminosity is calculated by assuming that the
electron-photon cross section is well described by the {\it electron
scattering} Thompson cross section. Here we discuss the modifications
to this simple picture by taking into account {\it\brem} opacity also:
due to the dependence of \brem emission on the {\it square} of the gas
density, it follows that the associated absorption cross section
increases {\it linearly} with the gas density, so that Eddington
luminosity is correspondingly reduced. It is then expected that this
opacity source will be important in the range spanned by {\it cold
accretion solutions} (see, Park \& Ostriker 1999). These high-density
conditions were certainly present at the epoch of galaxy formation,
thus a quantitative estimate of the reduction of Eddington luminosity
as a function of ambient gas density and temperature is of direct
interest when modeling the early stages of galaxy formation (see,
e.g., Haiman, Ciotti \& Ostriker 2003). In the following discussion we
often refer to formulae provided in Ciotti, Ostriker \& Pellegrini
(2003, hereafter COP): eq. (1) there will be indicated as COP1, and so
on.

\section{The effect of free-free absorption}

Let $\Lbh\equiv \eps\dot\Mbh c^2$ be the bolometric luminosity
associated with an accretion rate of $\dot\Mbh$, where $\eps$ is the
accretion efficiency and $c$ is the speed of light. The {\it momentum
transfer} to the gas electrons per unit time, frequency and volume at
the radius $r$ is given by (e.g., see Krolik 1999):
\begin{equation}
{\Delta p\over \Delta t}={\nscat (\nu)\over\Delta t}\times {h\nu\over c}=
{\Lbh (r, \nu)\sigtot (\nu)\nel (r)\over 4\pi r^2 c},
\end{equation}
where $\nscat (\nu)$ is the number of photon--electron interactions as
given in COP1, $\nel (r)$ is the electron number density, $h$ is the
Planck constant, and $\sigtot$ is the total electron cross--section
corresponding to the various absorption processes (e.g., see Krolik
1999). For simplicity we restrict to the {\it optically thin regime}:
as a consequence, from now on we assume
$\Lbh(r,\nu)\equiv\Lbh\times\fznu$ (see COP). The net acceleration
field experienced by the accreting gas with density profile $\rho(r)$
is given by $\gef (r)=-G\Mbh /r^2 +\int_0^{\infty}(\Delta p /\Delta
t)d\nu/\rho (r)$ where $G$ is the gravitational constant. By imposing
$\gef =0$ one obtains
\begin{equation}
\Ledd\equiv {\rho (r)\over\nel (r)} {4\pi G c\Mbh\over
            \int_0^{\infty}\sigtot(\nu)\fznu\,d\nu}:
\end{equation}
when $\sigtot =\sigT = 8\pi\qe^4/3\mel^2c^4\simeq 6.65\times 10^{-25}$
cm$^2$, the classical expression for $\Ledd$ is recovered. Here we
consider two different opacity sources for the emitted photons. The
first is due to {\it electron scattering} as described by the
Klein--Nishina cross section $\sigknnu =\sigT\tilde\sigkn(x)$, where
$x\equiv\nu /\nuT$ and $\nuT\equiv\mel c^2/h$ (see COP2).  The second
is due to {\it free--free} absorption: its cross cross section
$\sigff$ is obtained from the \brem emission formula (per unit volume,
frequency, over the solid angle, Spitzer 1978, hereafter S78)
\begin{equation}
\jff (\nu)={2^{11/2}(\pi /3)^{3/2}\nel\qe^6\sum_i\nei \Zi^2 g_{\rm ff,i}\over
           \mel^2 c^4}\left ({\mel c^2\over \kb T}\right)^{1/2}
           \exp{\left(-{h\nu\over\kb T}\right)}
\end{equation}
and the {\it Kirchhoff's law} $\jff(\nu)=\kappaff\Bnu (T)$, where
$\Bnu (T)=8\pi h\nu^3c^{-2}/[\exp{(h\nu/\kb T)} -1]$ and $\kb$ is the
Boltzmann constant. Accordingly,
\begin{equation}
\sigff (\nu)\equiv {\kappaff\over\nel}= \sigT 
       {\qe^2 c^2 <g_{\rm ff}>\sum_i\nei\Zi^2\over\sqrt{6\pi} h}
       \sqrt{{\mel c^2\over\kb T}}{1-\exp (-h\nu/\kb T)\over \nu^3},
\end{equation}
where $<g_{\rm ff}>$ is the mean {\it free-free Gaunt factor}.  We
assume that a mass fraction $X$ of the accreting material is made of
fully ionized hydrogen, and the remaining fraction $Y=1-X$ by fully
ionized helium. Thus, $\nh = X\rho/\mpr$, $\nhe = (1-X)\rho/4\mpr$,
$\nel=(1+X)\rho/2\mpr$, $\nt=(3+5X)\rho/4\mpr$, and finally
$\sum_{i=1}^2\nei\Zi^2=\rho /\mpr$. {\it In other words, at variance
with the standard electron scattering case, $\sigff$ depends
(linearly) on the gas density}.  As in COP, we also adopt a
(normalized) spectral distribution made by the sum of {\it two}
distinct contributions
\begin{equation}
\fznu={\fx+\rr\fuv\over 1 +\rr}.
\end{equation}
$\fx$ describes a (normalized) {\it double-slope}, non thermal
distribution of total luminosity $\Lx$, with low and high-frequency
slopes $\xi_1$ and $\xi_1 +\xi_2$, respectively, and break frequency
$\nub$ (COP9). $\fuv$ is a normalized {\it black--body} distribution
of temperature $\Tuv$ and total luminosity $\Luv\equiv\rr\Lx$ (COP10);
thus $\Lbh=\Lx+\Luv=(1+\rr)\Lx$.  Accordingly, we study
\begin{equation}
{\LeddKN\over\LeddKNff}=1+{
\int_0^{\infty}\sigff\fx d\nu +\rr\int_0^{\infty}\sigff\fuv d\nu
                          \over
\int_0^{\infty}\sigkn\fx d\nu+\rr\int_0^{\infty}\sigkn\fuv d\nu}.
\end{equation}
We now compute the four integrals appearing in eq. (6) for some
representative values of the parameters entering $\fznu$.  As in COP,
we fix $\xi_1=0.9$ and $\xi_2=0.7$, and $\rr\simeq 1-2$, so
reproducing the observed Compton temperature in quasar spectra (see,
e.g., Krolik 1999; Sazonov, Ostriker \& Sunyaev 2003). We obtain
\begin{equation}
\int_0^{\infty}\sigkn\fx d\nu\simeq 0.79\left
       ({\nub\over\nuT}\right )^{-0.05}\sigT
\end{equation}
as a very good fit over the range $0.1\leq \nub/\nuT\leq 10$. The
estimate of $\int_0^{\infty}\sigkn\fuv d\nu$ is trivial: in fact, the
99\% of $\Luv$ is emitted at $h\nu < 10\kb\Tuv \ll h\nuT$ for any
reasonable value of $\Tuv$. Thus $\sigkn\sim\sigT$, and from the
normalization condition
\begin{equation}
\int_0^{\infty}\sigkn\fuv d\nu\simeq\sigT.
\end{equation}
We then obtain
\begin{equation}
\int_0^{\infty}\sigff\fuv d\nu ={\rho\sigT\over\mpr}
{15 h^2\qe^2 c^2 <g_{\rm ff}>\over \sqrt{6}\pi^{9/2}\kb^3\Tuv^3}
\sqrt{{\mel c^2\over\kb T}}
\int_0^{\infty}{1-\exp (-\alpha x)\over \exp (x) -1}\,dx ,
\quad 
\end{equation}
where $\alpha\equiv \Tuv/T$, and the dimensionless integral equals
$\gamma +\Psi (1+\alpha)$, with $\gamma=0.5772...$ and $\Psi(x)\equiv
d\ln\Gamma (x)/dx$. Note that if $\Tuv=T$ then the integral equals 1;
moreover, $\gamma +\Psi (1+\alpha)\sim \pi^2\alpha/6$ for $\alpha\to
0$ and $\Psi (1+\alpha)\sim\ln (\alpha)$ when $\alpha\to\infty$. Thus
we have
\begin{equation}
\int_0^{\infty}\sigff\fuv d\nu\simeq {\rho\sigT\over\mpr}\times
                               \cases{
\displaystyle{{1.52\times 10^2\over \Tuv^2 T^{3/2}}},\quad
                                                    \Tuv\lsim T,\cr
\displaystyle{{0.93\times 10^2\over \Tuv^3 T^{1/2}}
              \left(\gamma +\ln{\Tuv\over T}\right )},\quad
                                                    \Tuv\gg T,} 
\end{equation}
where temperatures are in Kelvin, and we assumed $<g_{\rm ff}>\simeq
9.8$ (see S78).  The integral $\int_0^{\infty}\sigff\fx d\nu$ is
affected by a formal infrared divergency if (as in our case) $\fx\sim
\nu^{-\xi_1}$ with $\xi_1\geq -1$: in fact, $\sigff\sim\nu^{-2}$ for
$\nu\to 0$. In order to have a proper physical description we
integrate only at frequencies $\nu \geq \chi\nup$, where
\begin{equation}
\nup\equiv \sqrt{{\nel \qe^2\over \pi\mel}}\simeq 
6.35\times 10^3\sqrt{{(1+X)\rho\over\mpr}}\quad {\rm s}^{-1}
\end{equation}
is the plasma frequency. Under the condition $1\ll\chi\ll \kb
T/h\nup$, the opacity contribution comes mainly from the
Rayleigh-Jeans regime in eq. (4), and the asymptotic value of the
integral is
\begin{equation}
\int_{\chi\nup}^{\infty}\sigff\fx d\nu\sim
{7.8\times 10^{12}\over \chi^{1.9}(1+X)^{0.95} T^{3/2}}
\left({\nub\over\nuT}\right)^{-0.1}
\left({\rho\over\mpr}\right)^{0.05}\sigT .
\end{equation}
Thus, if $X=0.75$, $\nub/\nuT =0.5$, $\rr=2$ and $\Tuv=T$ we obtain
\begin{equation}
{\LeddKN\over\LeddKNff}\simeq 1+{1.7\times 10^{12}\over \chi^{1.9} T^{3/2}}
                                \left({\rho\over\mpr}\right)^{0.05}
+{50\over T^{7/2}}{\rho\over\mpr}.
\end{equation}
For $T\gsim 10^4$ K the free-free opacity contribution to Eddington
luminosity is fully dominated by the low frequency tail in the
non-thermal spectral distribution. For example, a reduction factor of
Eddington luminosity of the order of 10 is obtained for $\chi\simeq
300$ and $\chi\simeq 15$, when $T=10^5$ K and $T=10^6$ K, respectively.


Of course, the present analysis leaves open several questions, such as
the astrophysical status of the high-density gas at low temperatures,
the characteristic linear scale over which the gas becomes optically
thick, and the detailed shape of the non-thermal radiation
distribution in AGNs at (very) low frequencies. All these aspects of
the problem should be proprerly addressed when applying the present
results to numerical models.

\begin{theacknowledgments}
We thank Giuseppe Bertin, Bruno Coppi, and Bruce Draine for
enlightening comments; L.C. aknolwedges the warm hospitality of
Princeton University Observatory and the Institute of Astronomy at
Cambridge, where a large part of this work was carried out. L.C. was
partially supported by a grant CoFin 2000 of the Italian MIUR and by
the grant ASI I/R/105/00.

\end{theacknowledgments}

\bibliographystyle{aipprocl} 

{}

\end{document}